\documentclass[aps,prb,twocolumn,showpacs]{revtex4}
\usepackage{graphicx}

\begin{document}

\title{Hall effect and magnetoresistance in single crystals of  NdFeAsO$_{1-x}$F$_{x}$ }

\author{Peng Cheng, Huan Yang, Ying Jia, Lei Fang, Xiyu Zhu, Gang Mu, and Hai-Hu Wen}\email{hhwen@aphy.iphy.ac.cn }

\affiliation{National Laboratory for Superconductivity, Institute of
Physics and Beijing National Laboratory for Condensed Matter
Physics, Chinese Academy of Sciences, P. O. Box 603, Beijing 100190,
China}

\begin{abstract}
Hall effect and magnetoresistance have been measured on single
crystals of $NdFeAsO_{1-x}F_{x}$ with x = 0 ($T_c$ = 0 $\;$K) and x
= 0.18 ($T_c$ = 50 $\;$K). For the undoped samples, strong Hall
effect and magnetoresistance with strong temperature dependence were
found below about 150 K. The magnetoresistance was found to be as
large as 30\% at 15 K at a magnetic field of 9 T. From the transport
data we found that the transition near 155 K was accomplished in two
steps: first one occurs at 155 K which may be associated with the
structural transition, the second one takes place at about 140 K
which may correspond to the spin-density wave like transition. In
the superconducting sample with $T_c$ = 50 $\;$K, it is found that
the Hall coefficient also reveals a strong temperature dependence
with a negative sign. But the magnetoresistance becomes very weak
and does not satisfy the Kohler's scaling law. These dilemmatic
results (strong Hall effect and very weak magnetoresistance) prevent
to understand the normal state electric conduction by a simple
multi-band model by taking account the electron and hole pockets.
Detailed analysis further indicates that the strong temperature
dependence of $R_H$ cannot be easily understood with the simple
multi-band model either. A picture concerning a suppression to the
density of states at the Fermi energy in lowering temperature is
more reasonable. A comparison between the Hall coefficient of the
undoped sample and the superconducting sample suggests that the
doping may remove the nesting condition for the formation of the SDW
order, since both samples have very similar temperature dependence
above 175 K.
\end{abstract} 
\maketitle

\section{Introduction}

Since the discovery of superconductivity at 26$\;$K in iron-based
layered quaternary compound LaFeAsO$_{1-x}$F$_x$\cite{LaOFFeAs}, a
group of high temperature superconductors have been discovered. For
example, in electron doped region the highest critical temperature
$T_\mathrm{c}=55\;$K was found in
SmFeAsO$_{0.9}$F$_{0.1}$\cite{SmF_RZA}, while in the hole doped
region superconductivity at $T_\mathrm{c}=25\;$K was first found in
La$_{1-x}$Sr$_{x}$FeAsO \cite{LaSr_Wen} and later
$T_\mathrm{c}=38\;$K\cite{BaFeAs} in Ba$_{1-x}$K$_x$Fe$_2$As$_2$. At
the same time extensive efforts have been devoted to study the
nature of this new generation of high temperature superconductors.
Among them, very high upper critical field was infered through high
field measurements in these iron based
superconductors\cite{LaF_Hunte}, which indicates encouraging
potential applications. Low temperature specific heat\cite{LaF_Mu},
point contact tunneling spectrum\cite{LaF_Shan}, lower critical
field\cite{LaF_Ren}, and NMR \cite{NMR} etc. revealed unconventional
pairing symmetry in the superconducting state. Theoretical
calculations pointed out that the superconductivity in these
iron-based superconductors may emerge on several disconnected pieces
of the Fermi surface\cite{T1,T2,T3,T4,T5}, thus exhibiting multi-gap
effect. Many experiments have already shown that these new
superconductors exhibit multi-band
features\cite{LaF_Hunte,LaF_Zhu,LaF_Ren}. Spin-density-wave order
and structural distortion were observed in undoped LaOFeAs
system\cite{Daip,Nakai,Carlo}. All these experimental and
theoretical works indicate an unconventional superconducting
mechanism in this system. To get a deeper insight into the
superconducting mechanism, the normal state properties become very
essential to know. Unfortunately almost all the transport data so
far in the normal state were taken from polycrystalline samples,
this casts big doubts in drawing any solid conclusions. This
situation becomes very serious in the layered system, like the
iron-based superconductors. Recently our group has successfully
fabricated single crystals of NdFeAsO$_{1-x}$F$_x$ (x=0.18) with
$T_\mathrm{c}^\mathrm{onset}=50\;$K\cite{JiaY}, so it becomes
possible to measure the intrinsic transport properties of these
iron-based superconductors. Here for the first time we report the
measurements of Hall effect and magnetoresistance on
NdFeAsO$_{1-x}$F$_x$ (x=0, 0.18) single crystals. Our data provide
intrinsic and detailed information about the electron scattering in
the normal state.

\section{Experiment}

\begin{figure}
\includegraphics[width=8cm]{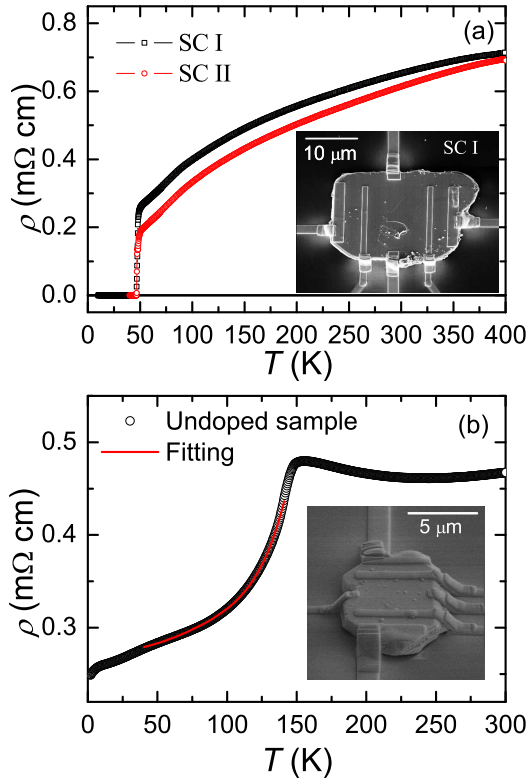}
\caption{(Color online) Temperature dependence of resistivity for
two superconducting (SC) NdFeAsO$_{0.82}$F$_{0.18}$ single crystals
(a) and one NdFeAsO single crystal (b). The insets show the scanning
electron microscope (SEM) picture of SC sample I and undoped sample
with the electric contacts of Pt metal made by Focused-Ion-Beam
technique for the resistance and Hall measurements. The solid line
in (b) shows the theoretical fitting by Eq.~\ref{SDWgap} which gives
the SDW energy gap about 1082 K. (see text)} \label{fig1}
\end{figure}

The crystals were made by flux method using NaCl as the flux, and
the detailed process of sample preparation was given elsewhere
\cite{JiaY}. The surface of the crystal looks rather flat. X-ray
diffraction pattern taken on one single crystal shows only (00l)
peaks and the Full-Width at the Half maximum (FWHM) of the (003)
peak is only 0.12$^\circ$, indicating good crystallinity. We made
electric contacts on them using the Pt deposition method of the
Focused-Ion-Beam (FIB) technique. As shown in the insets of
Fig.~\ref{fig1}, crystals with six Pt leads are shown and the
longitudinal and transverse resistance can be measured at the same
time. The measurements were carried out on a physical property
measurement system (PPMS, Quantum Design) with magnetic fields
perpendicular to the ab-plane of the samples and up to 9$\;$T. The
current density during the measurement was about
$400\sim500\;$A/cm$^2$. The in-plane longitudinal and the Hall
resistance were measured by either sweeping the magnetic field at a
fixed temperature or sweeping the temperature at a fixed magnetic
field.

\section{Experimental results and discussions}

In Fig.~\ref{fig1} (a), the resistive transitions of two
superconducting samples were shown. The superconducting (SC)
NdFeAsO$_{0.82}$F$_{0.18}$ samples have very sharp transitions,
i.e., the onset transition temperatures are about 50$\;$K and the
transition width is less than 2$\;$K (90\%$\rho_n$ and 1\%$\rho_n$),
which indicates good quality of the samples. Obviously, the general
shapes of the resistivity curves for the two samples do not show a
good linear metallic behavior but a strange curved feature persists
up to 400$\;$K. The resistivity at 55$\;$K is about 0.20 and
0.28$\;m\Omega\,$cm for the two samples respectively. The slight
difference between the two values may be attributed to the error in
the measurement on thickness by the FIB technique, or due to the
slight difference between the two samples. However, both values are
much smaller than 0.58$\;m\Omega\,$cm in a polycrystalline
sample\cite{NdF_RZA}. For the resistive curve of the undoped NdFeAsO
sample as shown in Fig.~\ref{fig1} (b), a strong resistivity anomaly
can be found near 150$\;$K. This anomaly was attributed to the
structural transition and/or the SDW formation. We will show below
that the onset point at about 150 K of the resistive transition may
be corresponding to the structural transition, and another one which
occurs at a lower temperature is related to the SDW
formation.\cite{Daip} For Fig.~\ref{fig1} (b), it should be noted
that the unusual decrease at about 6 K could be regarded as the
emergence of the ordering of the Nd ion. So the doping of F weakens
the spin density wave (SDW) and the structural phase transition, and
generates the superconductivity.

\subsection{MR and Hall effect of superconducting NdFeAsO$_{0.82}$F$_{0.18}$ crystals}

\begin{figure}
\includegraphics[width=8cm]{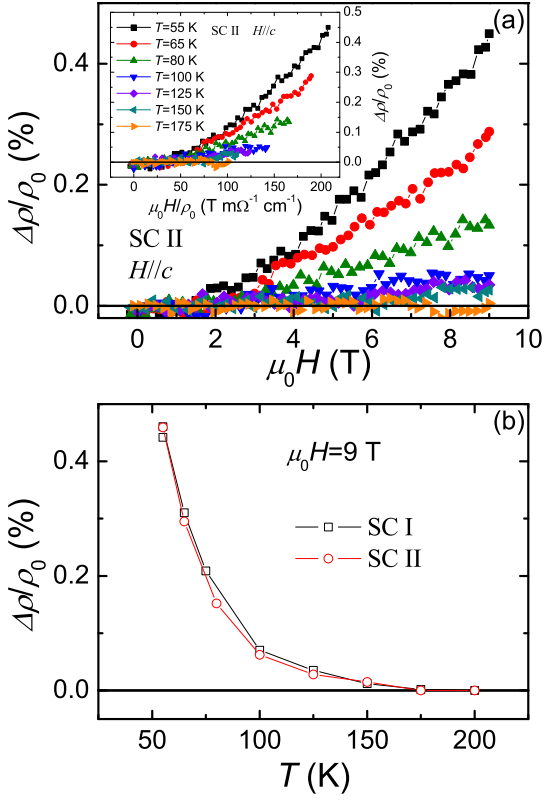}
\caption {(Color online) (a) Field dependence of MR
$\Delta\rho/\rho_0$ for SC sample II at different temperatures. The
inset shows the plot by following Kohler's law at different
temperatures, one can see that the Kohler's law is not obeyed. (b)
Temperature dependence of MR at 9$\;$T. One can see that the MR is
rather small and decreases rapidly with increasing temperature.
Above about 175$\;$K, the MR becomes negligible in our measurement
resolution. } \label{fig2}
\end{figure}

It is known that the magnetoresistance (MR) is a very powerful tool
to investigate the electronic scattering and the message of the
Fermi surface. For example, in MgB$_2$, a large magnetoresistance
was found which is closely related to the multiband
property\cite{LiQ,YangHall}. Fig.~\ref{fig2} (a) shows the field
dependence of MR for SC sample II. Here we define the
magnetoresistivity as $\Delta\rho=\rho(H)-\rho_0$, where $\rho(H)$
is the longitudinal resistivity at a magnetic field $H$ and $\rho_0$
is that at zero field. It can be observed that the MR is very weak,
i.e., less than 0.5\% at 55$\;$K, which is of the same magnitude as
the value in hole-doped cuprate
superconductors\cite{OngMR,MRLaSrCuO}. Fig.~\ref{fig2} (b) shows the
MR versus temperature at 9$\;$T. One can see that MR decays rapidly
with increasing temperature, and it cannot be detected in our
measuring resolution above 175$\;$K. Usually the MR effect may be
weakened by mixing the transport components with the magnetic field
along different directions of the crystallographic axes. However, in
our measurements the MR in the single crystal is one order of
magnitude smaller than the value in the polycrystalline samples
\cite{LaF_Zhu}. For many metallic materials with a symmetric Fermi
surface, the Kohler's law is normally obeyed. According to the
Kohler's law\cite{Kohler}, MR at different temperatures can be
scaled by the expression $\Delta\rho/\rho_0=f(H\tau)=F(H/\rho_0)$
with the assumption that scattering rate
$1/\tau(T)\propto\rho_0(T)$, here $f$ and $F$ represent some unknown
functions. For MgB$_2$, the Kohler's law is not obeyed because of
the multiband property\cite{LiQ}. Here we tried the scaling based on
the Kohler's law to the data from the single crystal (SC sample II),
the result is shown in the inset of Fig.2(a). Clearly, the MR data
measured at different temperatures do not overlap and the Kohler's
law is not obeyed in this material. This discrepancy may suggest
that in this new material there is multiband effect or the basic
assumption for the Kohler's law $1/\tau(T)\propto\rho_0(T)$ may not
be satisfied. Through the following analysis, we will see that the
latter may be more plausible.

\begin{figure}
\includegraphics[width=8cm]{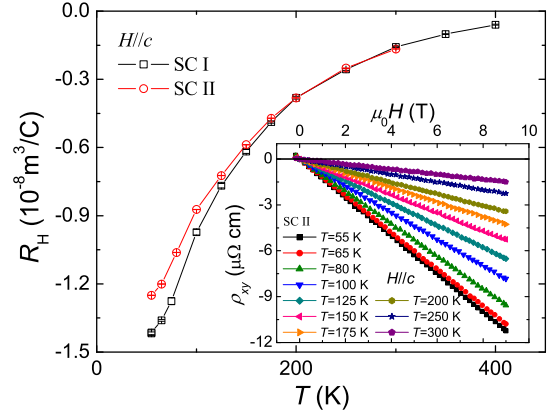}
\caption{(Color online) Temperature dependence of the Hall
coefficient $R_\mathrm{H}$ for the two superconducting samples.
Strong temperature dependence of Hall coefficient with a negative
sign can be seen obviously. The inset shows a good linear relation
between $\rho_{xy}$ and magnetic field at different temperatures for
Sample II.} \label{fig3}
\end{figure}

For a normal metal with Fermi liquid feature, the Hall coefficient
is constant versus temperature. However, the Hall coefficient varies
with temperature for a multiband material, such as MgB$_2$
\cite{YangHall}, or a sample with non-Fermi liquid behavior such as
the cuprate superconductors e.g. in Ref.~\cite{Ongreview}. In the
newly found iron based superconductors, $R_\mathrm{H}$ varies with
temperature in polycrystals\cite{LaF_WNL,LaF_Zhu}. People may
question that the temperature dependence of $R_H$ observed in
polycrystalline samples do not show an intrinsic property, since the
electric current flows randomly in different directions on different
grains, which gives rise to complexity in interpreting the Hall
effect. In the present work, the Hall resistance was carefully
measured on NdFeAsO$_{1-x}$F$_{x}$ single crystals, and this concern
can be removed completely. The inset of Fig.3 shows the raw data of
the transverse resistivity $\rho_{xy}$ at different temperatures,
which is in good linear relation against the magnetic field. In
Fig.~\ref{fig3}, the temperature dependence of the Hall coefficients
$R_\mathrm{H}=\rho_{xy}/\mu_0H$ were plotted, which decay
continuously with increasing temperature and behave in almost the
same way in the two samples. The value of $R_H$ at 400 K is about 20
times smaller than that at 55$\;$K.

\begin{figure}
\includegraphics[width=8cm]{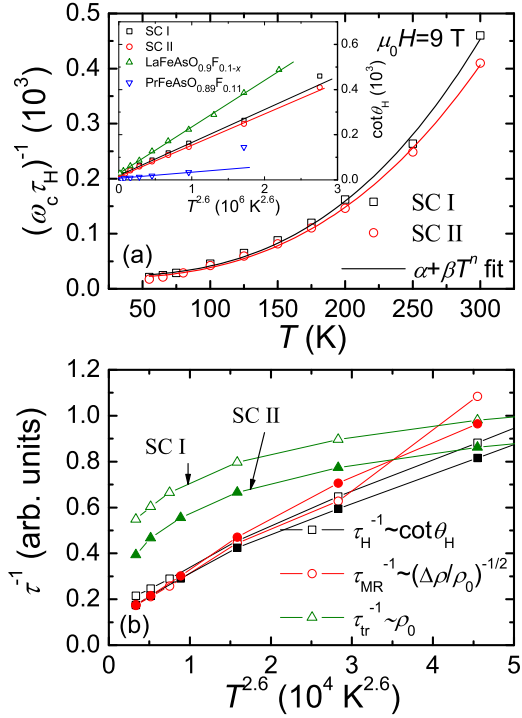}
\caption{(Color online) (a) Temperature dependence of
$(\omega\tau_\mathrm{H})^{-1}$ of the two samples. The solid lines
show the fit by the expression $\alpha+\beta T^n$ with
$n=2.82\pm0.11$ for SC sample I and $n=2.62\pm0.08$ for SC sample
II. The inset gives the linear behavior between
$\cot\theta_\mathrm{H}$ and $T^{2.6}$ for the single crystal and two
other  polycrystalline samples, and the solid lines shows the linear
fit to the experimental data (see text). (b) Temperature dependence
of $\tau_\mathrm{H}^{-1}$, $\tau_\mathrm{MR}^{-1}$, and
$\tau_\mathrm{tr}^{-1}$. } \label{fig4}
\end{figure}

A straightforward interpretation to the violation of the Kohler's
law and the strong temperature dependence of $R_H$ would be the
multiband effect, especially it is true when one assumes an almost
balanced contribution of electron and hole pockets. This however
contradicts the following facts. In a simple two band model, MR can
be expressed as follows by omitting the higher-order terms of
$B=\mu_0H$
\begin{equation}
\frac{\Delta\rho}{\rho_0}\simeq\frac{\sigma_1\sigma_2(\mu_1-\mu_2)^2B^2}{(\sigma_1+\sigma_2)^2},
\end{equation}
where $\sigma_i=n_ie^2\tau/m_i$ and $\mu_i=e\tau/m_i$ are
 the conductivity and the mobility of the $i^\mathrm{th}$
band respectively, with $n_i$ and $m_i$ meaning the charge carrier
density and the effective mass of the $i^{th}$ band, respectively.
The Hall coefficient can be expressed as
\begin{equation}
R_\mathrm{H}\simeq\frac{\sigma_1\mu_1+\sigma_2\mu_2}{(\sigma_1+\sigma_2)^2}.
\end{equation}
If assuming the sample that we investigate here has comparable
contributions from electron pocket and hole pocket, the
magnetoresistance should be quite strong as in $MgB_2$. This is
actually not the case. Band structure calculations\cite{T1,Fang,T4}
also show that when doping the parent phase $ReFeAsO$ with more than
10\% electrons like in our present case, the hole pockets will
shrink into small ones and the electron pockets expand a lot. In
this case it is very hard to believe that the hole pockets play
still an important role in the electric conduction. The two electron
pockets surrounding the $M-A$ are highly degenerate, therefore it is
natural to assume that they have very similar Fermi velocity and
electron mass. The electron scattering may also share high
similarities. It is thus very difficult to use eq.(2) to understand
the strong temperature dependence of $R_H$ since the scattering
times $\tau_1$ and $\tau_2$ are close to each other, and to the
first order assumption no more scattering time is involved in eq(2).
In order to check whether this is a more reasonable case we did the
calculation of
$\cot\theta_\mathrm{H}=\rho_{xx}/\rho_{xy}\equiv1/(\omega_\mathrm{c}\tau_\mathrm{H})$
which measures only the scattering rate $1/\tau$ for the
NdFeAsO$_{0.82}$F$_{0.18}$ single crystal (Here $\tau_H$ is the
$\tau$ determined from Hall angle, $\omega_\mathrm{c}$ is the
circling frequency). Since the temperature dependence of $\rho$ and
$R_\mathrm{H}$ are very complex and are difficult to be described by
the easy expression, we try to find out the relationship between
$\cot\theta_\mathrm{H}$ and $T$. Fig.~\ref{fig4} (a) shows the
temperature dependence of $\cot\theta_\mathrm{H}$. Then we try to
fit the result with the expression of $\alpha+\beta T^n$, and get
good fitting results with average exponent $n\sim2.7$. It is
worthwhile to note that the quantity
$\cot\theta_\mathrm{H}=\rho_{xx}/\rho_{xy}$ measures mainly the
scattering time and the information about the charge carrier density
is naturally separated away. This simple power-law like temperature
dependence of $\rho_{xx}/\rho_{xy}\propto1/\tau_H\propto\alpha+\beta
T^n$ is in sharp contrast with the temperature dependence of the
in-plane resistivity $\rho_{xx}$, manifesting strongly that the
charge carrier density may have a strong temperature dependence. We
also try to find the temperature dependence of
$\cot\theta_\mathrm{H}$ for other polycrystalline samples, and it is
interesting to see that the exponents from different samples are
quite close to each other: $n=2.64\pm0.07$ in
LaFeAsO$_{0.9}$F$_{0.1-x}$ below about 250$\;$K, and $n=2.57\pm0.11$
in PrFeAsO$_{0.89}$F$_{0.11}$ at temperatures below about 200$\;$K.
The inset of Fig.~\ref{fig4} (a) shows the Hall angle versus
$T^{2.6}$, and the good linear behavior can be observed with
different slopes. Therefore the relationship
$\cot\theta_\mathrm{H}=\alpha+\beta T^n$ (n=2-3) may be universal
for the iron based superconductors below some characteristic
temperatures.

\begin{figure}
\includegraphics[width=8cm]{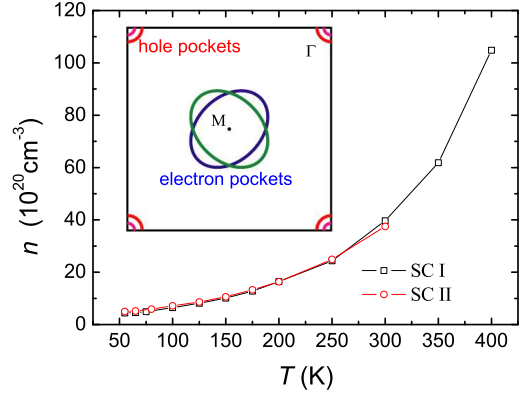}
\caption{(Color online) Temperature dependence of charge carrier
density $n=1/R_\mathrm{H}e$ of the two superconducting samples
calculated from $R_H$ based on the single band model. The inset
shows the schematic figure of the Fermi surface (see text) with a
heavy electron doping. } \label{fig5}
\end{figure}
Bear this idea in mind, we can actually understand the violation of
the Kohler's law very well. The basic requirement for the Kohler's
scaling law is $1/\tau(T)\propto\rho_0(T)$, this should not give
problems when dealing a metal with one band and a weak temperature
dependent density of states (DOS). However, when the DOS is changing
with temperature, the Kohler's law is certainly not followed.
Assuming that the two electron pockets are the major players of
electron conduction in our case, the electrons have slightly
different mass $m_1$ and $m_2$ on these two bands, but the charge
carrier density $n_1$ and $n_2$ and $\tau_1$ and $\tau_2$ are
roughly equal to each other, from eq.(1) we have
$\Delta\rho/\rho_0\simeq1/4(e^2\tau\Delta m/m)^2B^2$ with $\Delta m=
m_1-m_2$. Thus one expects that $(\Delta \rho/\rho_0)^{1/2}\propto
\tau$. In Fig.4 (b), we accumulate the $\tau$ values calculated in
different ways. First the $1/\tau_H$ calculated from the Hall angle
$\cot\theta_\mathrm{H}$ is proportional to $T^{2.6}$ as mentioned
before. This may tell the intrinsic message for the electron
scattering in the normal state. Then the $1/\tau_{tr}$ calculated
from resistivity exhibits a complicated structure which may reflect
a combined result of both the true scattering rate $1/\tau_H$ and
the temperature dependent DOS. It is found that the
magnetoresistance $(\Delta \rho/\rho_0)^{-1/2}\propto1/\tau_{MR}$
scales also with the power law $T^{2.6}$ which suggests $(\Delta
\rho/\rho_0)^{1/2}\propto \tau_H$. This consistent analysis reveal
that a more reasonable picture to understand the transport data on
the single crystals is to assume \emph{a depletion of DOS in
lowering temperature}, at least this is part of the reason for the
strong temperature dependence of the Hall coefficient $R_H$.

The depletion to DOS in lowering temperature is actually a common
feature in correlated materials. In cuprate superconductors, this
effect is associated with a term "pseudogap" in the normal state,
which has been observed by many tools. In the present iron-based
superconductors, pseudogap has actually been already inferred from
the NMR and photo-emission data\cite{NMR,ARPES,ZhouXJ}. In Fig.5,
based on on the single band model (assuming the two electron bands
are highly degenerate), we calculate the temperature dependence of
the charge carrier density $n=|1/eR_H|$. It should be mentioned that
the temperature dependence of $R_H$ in the iron-based system is much
stronger than that in the hole doped cuprate, where the most
accepted picture is the partial gapping of Fermi surface by the
pseudogap effect. In the iron-based superconductors, in order to
explain the very small magnetoresistance and the strong temperature
dependence of the Hall coefficient $R_H$, the picture concerning
partial gapping on the Fermi surface seems necessary.

\subsection{MR and Hall effect of undoped NdFeAsO}

The properties of the undoped samples are very different from the
superconducting ones, which can be found from the resistive curves
shown in Fig.~\ref{fig1}. A well accepted picture to understand the
resistivity anomaly at about 150 K is the formation of a SDW
gap\cite{Daip}. To analysis the data at temperatures below the SDW
transition, we tried the theory given by Anderson and Smith,
\cite{SDWRT}
\begin{equation}
\rho(T)=\rho_0+AT^2+BT(1+2T/\Delta)\exp(-\Delta/T).\label{SDWgap}
\end{equation}
Here, $\rho_0$ is the residual resistivity, $A$ and $B$ are the
fitting parameters; the $T^2$ term describes the Fermi liquid
behavior, while the last term describes the metallic ferromagnet or
antiferromagnet state (e.g., Ref.~\cite{Ugap}) with an energy gap
$\Delta$. The curve at temperatures from 40 K to 140 K can be well
fitted by the expression (see the solid line in Fig.1(b)), and an
SDW energy gap $\Delta\approx 1082\;$K is derived.

\begin{figure}
\includegraphics[width=8cm]{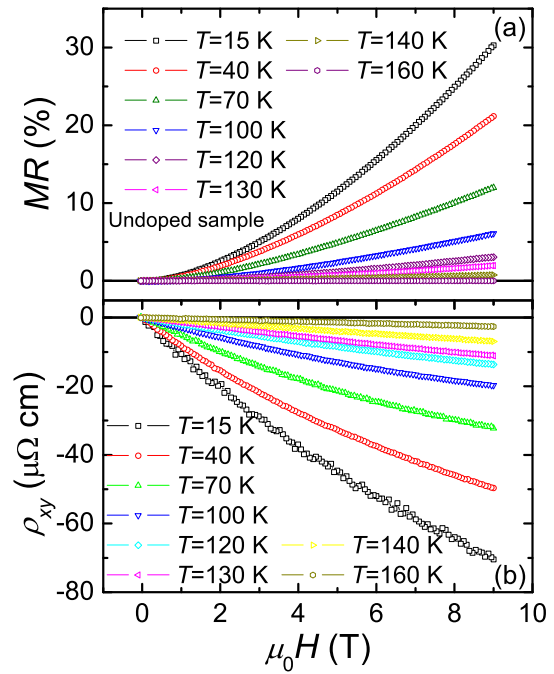}
\caption {(Color online) Field dependence of MR (a) and transverse
resistivity (b) for the undoped sample at different temperatures.
Large MR effect and a little nonlinear Hall effect can be found in
the undoped sample.} \label{fig6}
\end{figure}

\begin{figure}
\includegraphics[width=8cm]{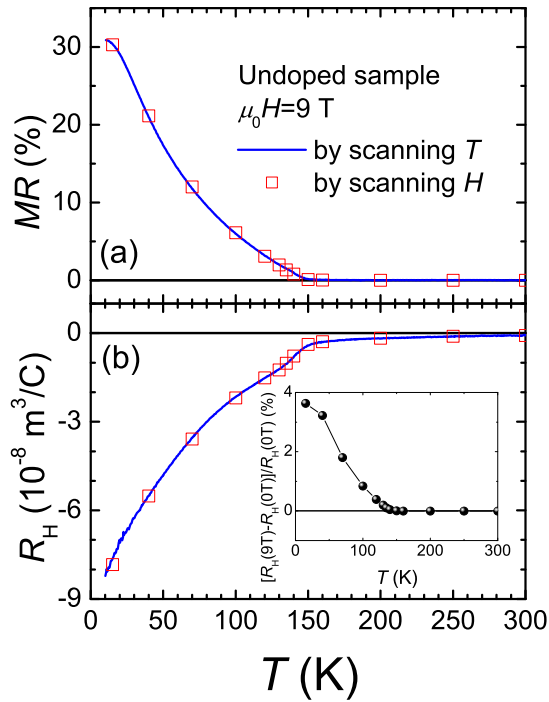}
\caption {(Color online) Temperature dependence of $MR$ (a) and Hall
coefficient $R_H$ at $\mu_0H=9\;T$. The inset of (b) shows the
magnitude of nonlinear Hall effect expressed by
$\left[R_H(9\mathrm{T})-R_H(0\mathrm{T})\right]/R_H(0\mathrm{T})$.}
\label{fig7}
\end{figure}

\begin{figure}
\includegraphics[width=8cm]{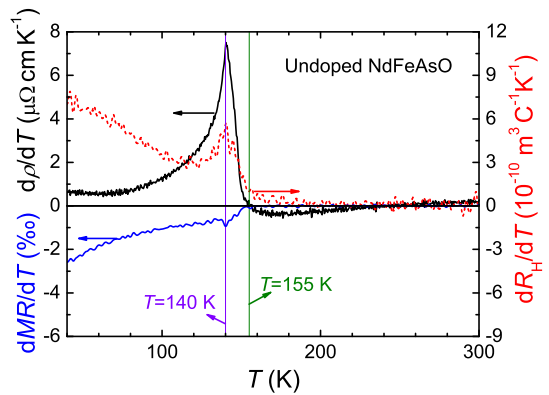}
\caption {(Color online) Differential of $\rho$ [from
Fig.~\ref{fig1}(b)], $MR$ [from Fig.~\ref{fig7}(a)], and $R_H$ [from
Fig.~\ref{fig7}(b)] versus $T$. Two anomalies could be seen at the
temperatures of about 140 K and 155 K, which may correspond to the
SDW and the structural phase transitions, respectively.}
\label{fig8}
\end{figure}
The SDW state is very interesting and worth further investigation.
In Fig.~\ref{fig6} the field dependence of MR and transverse
resistivity were shown. The MR effect is very strong at low
temperatures, and the magnitude is as large as 30\% at 9 T and 15 K.
In Fig.~\ref{fig6} (b), the transverse resistivity shows nonlinear
behavior at low temperatures. Both of them show very different
behaviors from the superconducting samples. Fig.~\ref{fig7} shows
the temperature dependence of $MR$, $R_H$, and the magnitude of
nonlinear Hall effect. Obviously, the curves measured by sweeping
temperature at fixed field are consistent with the curves measured
by sweeping the field at fixed temperatures, as shown in
Fig.~\ref{fig7}. All of the three curves have a clear kink at about
155 K, i.e., at the temperatures below it, both of the values and
the changing rates are enhanced gradually, while above that point
the values become very small (even vanished) and change slowly with
temperature. The large MR and nonlinear Hall effect may be
associated with the multiband effect or the complex scattering
between the itinerant electrons and the spin moment. The magnetic
field will make the long range spin order (the SDW here) frustrated
which leads to more strong scattering to the electrons and thus
larger resistivity. In this material, we cannot give a definite
judgement about the origin of the strong magnetoresistance: whether
it is due to the multiband effect or due to the scattering with
magnetic moments. From the temperature dependence of Hall
coefficient $R_H$, one can already see that the transition is
accomplished in two steps. A first inflecting point occurs at about
155 K, then it increases in magnitude and another kink appears at a
lower temperature, ca. 140 K. The transition at about 155 K may
correspond to the structural phase transition, while the lower one
at about 140 K may relate to the SDW formation.\cite{NdSDW,Daip} In
order to have a close scrutiny on these two transitions, the
differential of the $\rho$-$T$, $MR$-$T$, and $R_H$-$T$ curves are
shown in Fig.~\ref{fig8}. Now it becomes very clear that there are
indeed two characteristic points at temperatures of around 140 K and
155 K, respectively. So the anomalous behaviors of $d\rho/dT$-$T$,
$dMR/dT$-$T$, and $dR_H/dT$-$T$ may start from the structural phase
transition, and show a cusp-like behavior at the SDW transition
temperature. Hopefully a future investigation will give an
explanation to this close correspondence.

\begin{figure}
\includegraphics[width=8cm]{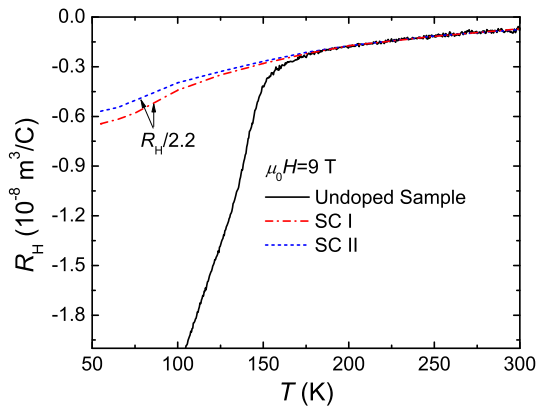}
\caption {(Color online) Temperature dependence of $R_H$ of the
undoped sample and $R_H/2.2$ of the two superconducting samples. The
curves at high temperature above 175 K overlap very well, which
suggests that the properties of the samples are very similar to each
other in that temperature region. } \label{fig9}
\end{figure}

\subsection{Comparison between the undoped $NdFeAsO$ and the superconducting $NdFeAsO_{0.82}F_{0.18}$}
From the results mentioned above, the F doping weakens the SDW and
the structural phase transition and finally generates
superconductivity beyond a certain doping level. However, the
situation at high temperatures above the two transitions are still
unknown. It would be very interesting to have a comparison between
properties of these two very different samples. For that the Hall
coefficient of the undoped sample and $R_H/2.2$ for two
superconducting samples are shown in Fig.~\ref{fig9}. Although the
absolute value of $R_H$ of the superconducting sample is about two
 times larger than that of the undoped sample, it is surprising to see that
the $R_H$ for two samples can be nicely scaled with a simple
multiplication of 2.2 for $R_H$ of the undoped sample from 175 K all
the way up to 300 K. This suggests that both the undoped
non-superconducting samples and the doped superconducting samples
may have quite similar Fermi surfaces with different charge carrier
densities in high temperature region. This gives support to the
theoretical calculation that the charge doping, may not shift the
Fermi energy too much, but rather lift off the very condition for
the formation of the SDW.\cite{Singh} This is a very important
conclusion which suggests that the rigid band model is certainly
insufficient to explain the doping effect. On the other hand, one
can see that the temperature dependence of resistivity are very
different for the two samples in the same temperature region,
indicating a very different electron scattering mechanism. A deeper
insight to this issue is very interesting and will be carried out in
the future investigation.

\section{Summary}

In summary, for the first time the MR and Hall effect have been
investigated on single crystals of superconducting
NdFeAsO$_{0.82}$F$_{0.18}$ samples and undoped non-superconducting
NdFeAsO ones. A strong temperature dependence of the Hall
coefficient $R_H$ has been found up to 400 K for both samples
although a greatly enhanced $R_H$ was observed below about 155 K for
the undoped one. It is further found that the MR for the
superconducting sample is very weak in the normal state and does not
obey the Kohler's scaling law. A picture concerning the partial
gapping to the density of states at the Fermi surface can
consistently explain the dilemmatic phenomenon: strong Hall effect
(as well as strong temperature dependence) but very weak
magnetoresistance the superconducting sample. For the undoped
samples, however, a very strong magnetoresistance was observed in
low temperature region, which could be due to the multiband effect
or the electronic scattering with the magnetic moment. The SDW gap
about 1082$\;$K for undoped NdFeAsO is derived from the $\rho$-$T$
curve for the first time. Two step transitions are observed from the
transport properties including $\rho(T)$, MR and $R_H(T)$ which
change their behaviors at about 155 K and 140 K. These two
transitions are in accord very well with the structural phase
transition and the SDW formation. The Hall coefficient $R_H$ for
both undoped and superconducting samples at temperatures above
175$\;$K show very similar behavior, which suggests the similar
Fermi surfaces for both systems and the inapplicability of the rigid
band model in explaining the doping effect.

\section{Acknowledgments}

We are grateful to Tao Xiang, Zhong Fang, Xi Dai and Zidan Wang for
fruitful discussions. This work is supported by the Natural Science
Foundation of China, the Ministry of Science and Technology of China
(973 project: 2006CB01000, 2006CB921802), the Knowledge Innovation
Project of Chinese Academy of Sciences (ITSNEM).

\end{document}